\documentclass{article}

\usepackage[english]{babel}
\usepackage[square,numbers,compress,sort]{natbib}
\usepackage{amsmath}
\usepackage{graphicx,subcaption}
\usepackage{mathtools}
\usepackage{gensymb} 
\usepackage{authblk}
\usepackage{tikz}
\usepackage{pgfplots}
\usepackage{csvsimple}
\usepackage{pgfplotstable}
\usepackage{xcolor}
\usepackage{rotating}
\usepackage[normalem]{ulem}
\usepackage[colorlinks=true,linkcolor=blue,urlcolor=cyan,citecolor=red,anchorcolor=blue]{hyperref}
\usetikzlibrary{pgfplots.patchplots}
\usetikzlibrary{pgfplots.dateplot}
\usetikzlibrary{pgfplots.colormaps}
\usetikzlibrary{pgfplots.groupplots}
\usetikzlibrary{pgfplots.polar}
\usetikzlibrary{pgfplots.units}
\usepgfplotslibrary{fillbetween}
\usepgfplotslibrary{groupplots}

\DeclarePairedDelimiter\ket{\lvert}{\rangle}
\DeclarePairedDelimiterX\braket[2]{\langle}{\rangle}{#1\,\delimsize\vert\,\mathopen{}#2}

\title{Apparatus for quantum-mixture research in microgravity}

\author[1,2]{Baptist Piest\thanks{Correspondence to: rasel@iqo.uni-hannover.de and baptist.piest@obspm.fr}}
\author[1]{Jonas Böhm}
\author[1,3]{Timothé Estrampes}
\author[1]{Priyanka Guggilam}
\author[1]{Annie Pichery}

\author[4]{Pawe\l{} Arciszewski}
\author[1]{Wolfgang Bartosch}
\author[5]{Sören Boles}
\author[4]{Klaus Döringshoff}
\author[6]{Michael Elsen}

\author[7]{Ortwin Hellmig}
\author[8]{Christian Kürbis}
\author[1]{Dorthe Leopoldt}
\author[1]{Gabriel Müller}
\author[1]{Alexandros Papakonstantinou}
\author[1]{Christian Reichelt}
\author[5]{André Wenzlawski}
\author[1]{Thijs Wendrich}
\author[3]{\'Eric Charron}
\author[9]{Christoph Lotz}
\author[4]{Achim Peters}
\author[7]{Klaus Sengstock}
\author[8]{Andreas Wicht}
\author[5]{Patrick Windpassinger}
\author[6]{Jens Grosse}
\author[1]{Naceur Gaaloul}
\author[1]{Ernst Maria Rasel\textsuperscript{*}}

\affil[1]{Institut f\"ur Quantenoptik, Leibniz Universit\"at Hannover, Welfengarten 1, 30167 Hannover, Germany}
\affil[2]{LTE, Observatoire de Paris, Université PSL, CNRS, Sorbonne Université, 61 avenue de l’Observatoire, 75014 Paris, France}
\affil[3]{Universit\'e Paris-Saclay, CNRS, Institut des Sciences Mol\'eculaires d'Orsay, 91405 Orsay, France}
\affil[4]{Institut f\"ur Physik, Humboldt-Universit\"at zu Berlin, Newtonstra{\ss}e 15, 12489 Berlin, Germany}
\affil[5]{Institut f\"ur Physik, Johannes Gutenberg-Universit\"at Mainz, 55099 Mainz, Germany}
\affil[6]{Zentrum f\"ur angewandte Raumfahrttechnologie und Mikrogravitation, Universit\"at Bremen, Am Fallturm 2, 28359 Bremen, Germany}
\affil[7]{Institut für Quantenphysik, Universität Hamburg, Luruper Chaussee 149, 22761 Hamburg, Germany}
\affil[8]{Ferdinand-Braun-Institut (FBH), Gustav-Kirchhof-Str. 4, 12489 Berlin, Germany}
\affil[9]{Institut f\"ur Transport und Automatisierungstechnik (ITA), Leibniz Universit\"at Hannover, Callinstraße 36, 30167 Hannover, Germany} 

\begin{document}

\maketitle
\thispagestyle{empty}
\enlargethispage{3\baselineskip}

\begin{abstract}
Experiments with ultracold quantum gases are a rapidly advancing research field with many applications in fundamental physics and quantum technology. Here, we report on a high-flux generation of Bose-Einstein condensate mixtures of $^{41}$K and $^{87}$Rb, using a fully integrated sounding rocket setup. 
We compare the release and the free expansion of the quantum mixtures obtained with the apparatus placed on ground or in free fall in an Einstein-Elevator. The release dynamics are governed by the intra- and interspecies interactions as well as the decaying magnetic field during the release. The latter can be minimized by a dedicated switch-off protocol of the trap generating currents where an exact model enabled us to characterize the interaction effects. 
Our results establish a new benchmark for generating ultracold mixtures on mobile platforms, with direct relevance for future experiments on interacting quantum gases and tests of the equivalence principle in space. 

\end{abstract}

\section{Introduction}
Microgravity opens up new perspectives for experiments exploring mixtures of interacting quantum gases \cite{Baroni2024}. These experiments range from the enhanced generation of Feshbach molecules \cite{Dincao_2017,DIncao2022}, interaction-driven generation of mixed species quantum bubbles \cite{Wolf2022, Lundblad2023} to high-precision tests of the Einstein equivalence principle and the search for new forces \cite{Ahlers2022}.
Microgravity experiments are either based on atom chips \cite{Muentinga2013, rudolph2015high, Becker2018, Elliott2018, Elliott2023} or all-optical \cite{Condon2019PRL, Vogt2020PRA, He2023, pelluet2024} generation of quantum gases and mixtures. Due to their low power consumption, small size and high robustness, atom chip-based setups are the baseline of future space-borne setups such as BECCAL \cite{Frye2021}, CARIOQA \cite{leveque2022carioqa}, CAI \cite{Trimeche2019}, QGG \cite{Stray2025} or STE-QUEST \cite{Ahlers2022}. Magnetic lensing techniques led to picokelvin expansion energies in space-borne atom chip single-species experiments \cite{Deppner2021, Gaaloul2022}. While recent advancements have led to the creation of atomic shell potentials \cite{Carollo2022}, atom interferometry \cite{Lachmann2021NatComms, Williams2024}, and magnetometry \cite{meister2025spacemagnetometrydifferentialatom}, as well as the first observation of interacting ultracold mixtures in space \cite{Elliott2023}, current microgravity-based BEC experiments fall short in the required atomic flux necessary for finding new Physics such as tests of the equivalence principle \cite{Ahlers2022} or gravitational wave detection and dark matter search \cite{ElNeaj2020}.
Another challenge in atom chip-based setups is that magnetic forces during and after release can cause side effects, which may obscure the actual measurement and complicate the data analysis \cite{Gaaloul2022}.
In particular, differential release velocities have been shown to induce phase shifts that can mimic violations of the equivalence principle in interferometric tests, and are recognized as one of the leading systematic errors in high-precision experiments \cite{Asenbaum2020, Loriani2019}. Cancellation techniques have been suggested and successfully implemented experimentally but require precise knowledge of the differential velocities \cite{Roura2014, dAmico2017, Overstreet2018}.\\
Here, we present the high-flux generation of a quantum degenerate K-Rb mixture in an atom chip apparatus suitable for space operation, surpassing current state-of-the-art realizations by an order of magnitude in atomic flux \cite{Elliott2018, Elliott2023}. Moreover, we control the differential trajectory of both atom species after release where the dynamics are governed by the interactions as well as the time-dependent release mechanism. We study the release of the atoms on ground and in microgravity, provided by the Einstein-Elevator at Leibniz-University Hannover (LUH) \cite{Lotz2017}. We apply a realistic theoretical framework describing the 3-dimensional evolution of interacting quantum mixtures \cite{Pichery2023} and achieve a high level of agreement with the experimental results. 
On ground, we can reliably generate quantum mixtures at various inclination angles of the experimental setup, offering unique insights into the behaviour of interacting mixtures shaped by quantum repulsion in conjunction with changing gravity conditions. These findings are confronted with the ones obtained in the absence of the gravitational sag and the dominance of atomic interactions.
These results demonstrate the potential of atom chip setups to perform metrological studies requiring high atomic mixture flux \cite{Ahlers2022, Bassi2022} as well as for proposed microgravity experiments that rely on the collocation of atomic mixtures \cite{Wolf2022, Lundblad2023, DIncao2022, Dincao_2017}.

\section{Results}

\subsection{High-flux generation of $^{41}$K and $^{87}$Rb BECs on an atom chip}
\label{sec:generation}
\begin{figure}
    \centering
    \includegraphics[width=1.0\textwidth]{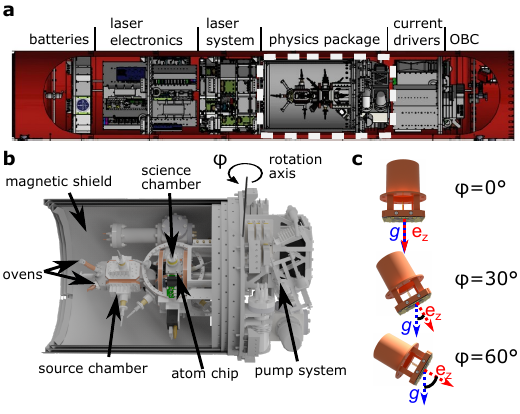}
    \caption{Overview of the experimental setup. a) Overall payload. The physics package is highlighted with a white dashed box. OBC: on-board computer. b) 3D drawing of the MAIUS-B physics package. When fixed in a suitable mounting frame, the module is rotatable around the indicated rotation axis, here shown for $\varphi=0^\circ$. c) Orientation of the atom chip inside the vacuum chamber for three different angles $\varphi$ between $z$-axis and gravity direction $g$. Fig. 1a) is adapted from \cite{Elsen2023}, licensed under CC BY 4.0. Modifications include revised annotations and rotation of image.}
    \label{fig:setup}
\end{figure}

The described experiments were performed on the MAIUS-B sounding rocket payload in different configurations \cite{Elsen2023}. The payload has been thoroughly qualified for a sounding rocket flight carried out on the 2nd December 2023 at ESRANGE.\\
The experimental setup is shown in Fig.~\ref{fig:setup}. For extended use in the laboratory, the payload can be continuously operated with water cooling, external power supplies and control units. For the microgravity results presented in this study, the modules are distributed on an experimental carrier to be compatible with the Einstein-Elevator at LUH \cite{Lotz2017, Raudonis2023}. An experimental cycle starts by simultaneously loading $^{87}$Rb and $^{41}$K atoms from a 2D-magneto-optical trap (2D-MOT) into a 3D-MOT. The quadrupole field of the 3D-MOT is generated by external Helmholtz coils and a mesoscopic U-structure of the three-layer atom chip. The three layers consist of a mesoscopic chip (U-structure, H-structure), a base chip (Z-structure) and a science chip (Z-structure) \cite{Piest2021}. Laser light addressing the D2-lines of $^{41}$K (767.7 nm) and $^{87}$Rb (780.2 nm) for cooling, optical pumping and absorption detection is generated by seven distinct external cavity diode lasers and guided to the collimators by optical fibers \cite{Kuerbis2020, Elsen2023}. The duration of the $^{41}$K- and $^{87}$Rb-MOTs can be tuned independently by switching the lasers with acousto-optical modulators. This allows to tune the ratio between the two species in the final condensates. A fixed loading time of 500$\,$ms for the $^{41}$K-MOT with an adjustable duration of the $^{87}$Rb-MOT turned out to be a suitable choice for the generation of mixtures with tunable ratios. We emphasize that it is not decisive to use different MOT loading times and it is also possible to achieve tunable mixture ratios by other methods, for example by changing the loading rate using different cooling frequencies of the 2D-MOT lasers. 
As discussed in Section~\ref{sec:losses}, higher numbers of trapped $^{87}$Rb atoms are possible but would prevent the generation of $^{41}$K condensates due to inelastic collisions between the co-trapped species.

\begin{figure}
    \centering
    \includegraphics[width=0.8\textwidth]{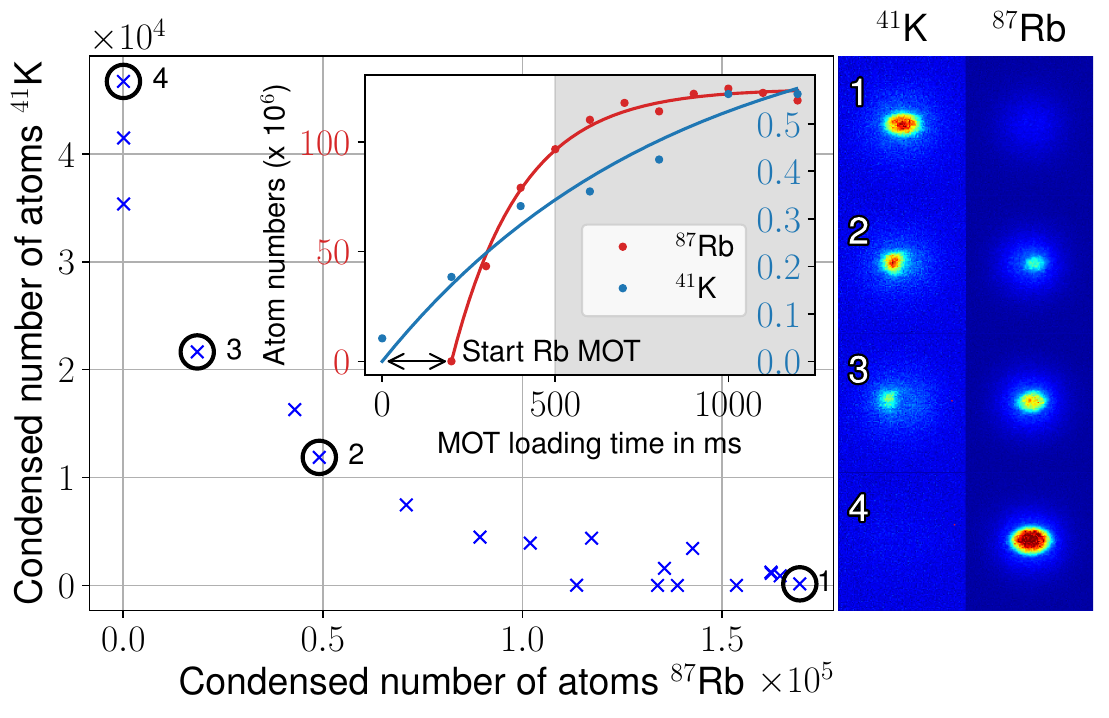}
    \caption{Generation of BEC mixtures with different ratios in the BEC fraction of the two species on ground. Corresponding absorption images for four different configurations are shown exemplarily on the right. The images are normalized for each species separately for better visibility. The inset shows typical loading curves of the magnetic traps. In this example, we load the K-MOT for 500$\,$ms and the Rb-MOT for 300$\,$ms. The possible magnetic trap loading curves for longer times is highlighted in grey.}
    \label{fig:KRb_Evap}
\end{figure}

Following the combined 3D-MOT stage, the atomic ensembles undergo spatial compression and additional cooling over $11\,\mathrm{ms}$ by adjusting the positions of the compressed MOTs (CMOTs) to maximize atoms loaded into the magnetic trap. By subsequent optical molasses cooling, we are able to reduce the temperatures to $43\,\mu$K for $2.6 \cdot 10^{7}$ $^{41}$K atoms and $10\,\mu$K for $9.4 \cdot 10^{8}$ $^{87}$Rb atoms, respectively. To reach sub-Doppler temperatures in the $^{41}$K molasses, we ramp the detuning of the cooling laser from $\delta=-3\,$MHz to $\delta=-8\,$MHz while switching the repumper to very low intensities $I_{rep}\approx I_{cool}/100$ with the cooling laser intensity $I_{cool}$ \cite{Landini2011}.\\ 
Before trapping the atoms magnetically, we apply a 1$\,$ms pulse of circularly polarized laser light resonant to the transition $\ket{F=2} \longleftrightarrow \ket{F'=2}$ for both species. This leads to an accumulation of atoms in the magnetically trappable state $\ket{F,m_f}=\ket{2,2}$.
By switching on the mesoscopic H-structure of the atom chip together with a Z-structure of the base chip, we load up to $7\cdot10^6$ ($^{41}$K) and $5\cdot10^8$ ($^{87}$Rb) atoms into a large volume magnetic trap where atoms cover an effective volume equivalent of Gaussian widths of axially $\sigma_x \approx 1200\,$µm and radially $\sigma_r \approx 530\,$µm at a temperature of approx. 85\,µK. This magnetic trap is loaded into a harmonic cylindrical trap close to the atom chip, generated by the Z-structures of the base and science chip \cite{Piest2021}. The intermediate step of using a large volume magnetic trap to mediate the size mismatch of the CMOT to the microscopic chip trap has previously been shown in drop tower experiments \cite{rudolph2015high} and is essential to achieve the reported flux.
Due to the high trapping frequencies, with axial frequency $f_{x,\mathrm{Rb}} = 23\,$Hz and radial frequency $f_{r,\mathrm{Rb}} = 910\,$Hz for $^{87}$Rb, and axial frequency $f_{x,\mathrm{K}} = 33\,$Hz and radial frequency $f_{r,\mathrm{K}} = 1326\,$Hz for $^{41}$K, this trap permits efficient evaporative cooling of $^{87}$Rb and sympathetic cooling of $^{41}$K, benefiting from high thermalization rates \cite{Delannoy2001}.\\
With the atom chip, we deploy two microwave fields resonant to the $\ket{F,m_f}=\ket{2,2} \longleftrightarrow \ket{1,1}$ and $\ket{2,1} \longleftrightarrow \ket{1,1}$ transitions of $^{87}$Rb. We use microwave fields instead of a single radiofrequency to prevent the simultaneous evaporation of the $^{41}$K atoms. The former microwave is used for evaporation while the latter is used to continuously depump remaining $^{87}$Rb-atoms trapped in the $\ket{2,1}$ state. These impurity states are immune to the evaporation microwave at later evaporation stages and would prevent the formation of a $^{87}$Rb BEC due to the additional thermal load. Additionally, as detailed in the methods section, $^{87}$Rb-atoms in $\ket{2,1}$ are highly detrimental to the co-trapped $^{41}$K-atoms due to inelastic collisions. Microwave depumping is a widely used technique to limit interspecies collisional losses \cite{Elliott2018, Modugno2001, KlemptPhD, Campbell2010, Haas2007, Marzok2007, Wang2011}. The co-trapped $^{41}$K-atoms are cooled sympathetically by fast thermalization with the $^{87}$Rb-atoms. The trajectory of the evaporation microwave consists of 9 concatenated ramps and an intermediate decompression step. The decompression is achieved by reducing the bias field from  $28.8\,$G to $20.3\,$G and reduces losses due to three-body collisions between the $^{87}$Rb atoms.The decompressed trap has trap frequencies of $f_{x,\mathrm{Rb}} = 25\,$Hz and $f_{r, \mathrm{Rb}}=397\,$Hz for $^{87}$Rb, and $f_{x,\mathrm{K}} = 36\,$Hz and $f_{r, \mathrm{K}}=578\,$Hz for $^{41}$K.
\begin{figure}
\includegraphics[width=\linewidth]{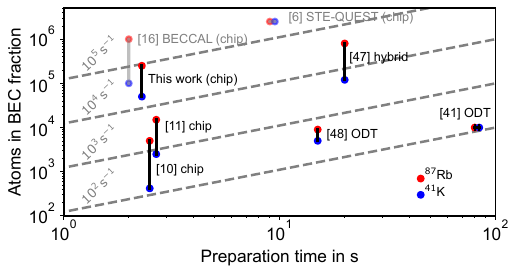}
\caption{Comparison of atom numbers and preparation times of BECs of $^{87}$Rb and $^{41}$K in various experiments. The red (blue) dots represent the atom number in the BEC fractions of $^{87}$Rb ($^{41}$K). The dotted lines indicate the constant atomic flux (numbers on the left side). The required capabilities of the future space-borne experiments BECCAL and STE-QUEST are also plotted. Note that the data points show the atom numbers for the single-species-optimized operation except for STE-QUEST, where they are targeted in mixture operation. Sources: \cite{Burchianti2018,Elliott2018,Thalhammer2008,Modugno2001,Elliott2023,Frye2021,Ahlers2022}.}
\label{fig:comparison}
\end{figure}
For the mixture generation, only the final frequency of the last ramp is increased by (70$\pm$10) kHz compared to the $^{87}$Rb-only case. This adjustment compensates for the small spatial shift of a few micrometers caused by the repulsive interaction with the cold $^{41}$K atoms, which leads to an energy shift of the $^{87}$Rb atoms. In microgravity, no change of the final evaporation frequency was applied between the single and mixed BEC cases.\\
After $1.7\,$s of evaporation, we generate BEC mixtures with a tuning range between $1.5\cdot10^5$ $^{87}$Rb-atoms and $5.0\cdot10^4$ $^{41}$K-atoms. By changing the duration of the $^{87}$Rb MOT, different mixture ratios can be achieved (Fig.~\ref{fig:KRb_Evap}). The trend of the curve demonstrates that an increased number of $^{87}$Rb atoms leads to a decrease of condensed $^{41}$K atoms. In a single species optimized sequence, we generate up to $2.5\cdot 10^5$ atoms in a pure $^{87}$Rb BEC. The total duration of one experimental cycle amounts to 2.3$\,$s.
Fig.~\ref{fig:comparison} shows the comparison of the performance of the apparatus to other BEC experiments working with the same atomic species. Our approach of combining a 2D-MOT and a three-layer atom chip - used for both MOT loading and evaporative cooling - allows us to achieve a high BEC mixture flux in a compact setup.
It surpasses the performance of other mobile and compact setups by an order of magnitude \cite{Elliott2018, Elliott2023}, and sets the stage for the upcoming generation of atom chip-based space-borne mixture experiments \cite{Frye2021,Ahlers2022}.

\subsection{Release dynamics of atom chip-based traps} \label{sec:releasedynamics}

Trapped atoms are commonly released into free fall by turning off the confining potentials created by a Z-shaped wire and an external bias field.
Our magnetic Ioffe-Pritchard trap is formed by $I_{\textrm{SC-z}}=2\,$A on the science chip and $I_{\textrm{BC-z}}=5.12\,$A on the base chip combined with a bias field of $B_y=20.3\,$G and an offset field of $B_x=-1.3\,$G to lift the trap bottom and prevent Majorana losses.

We observe that a naive switch-off, reducing the chip currents to zero while keeping the bias field on, results in a kick of the atoms towards the atom chip. Despite their low inductance, atom chip wires have typical switching times of a few $10\,$\textmu s. These short transients, along with additional eddy currents induced in the copper mount of the atom chip holder and the titanium vacuum chamber, generate time-dependent magnetic gradients which lead to a finite release velocity of the atoms although they were at rest in the trap.

This has low effect in most ground-based single species experiments as the resulting kick is much smaller than the gravity induced motion for typical free fall times. 

However, in mixture experiments, time-dependent magnetic gradients lead to a differential acceleration $\Delta a$ which separates both atomic species with masses $m_\mathrm{Rb}$ and $m_\mathrm{K}$ and equal magnetic moment $\boldsymbol{\mu}$ according to
\begin{equation}
    \label{eqn:differential_acceleration}
    \Delta \boldsymbol{a}(t) = \boldsymbol{a}_\mathrm{K}(t) - \boldsymbol{a}_\mathrm{Rb}(t) = \frac{m_\mathrm{K} - m _\mathrm{Rb}}{m_\mathrm{K} \, m _\mathrm{Rb}} \, \nabla \left( \boldsymbol{\mu} \cdot \boldsymbol{B}(\boldsymbol{r},t) \right).
\end{equation}
In case of a vanishing magnetic field gradient, there is no differential acceleration while, for a non-vanishing gradient at any time $t$ during the switch-off, we obtain different trajectories for each species. Velocity and position differences between species lead to systematic errors in atom interferometry \cite{PhysRevD.102.124043,Loriani2019, Struckmann2024} and limit applications requiring precise co-location, such as shell-shaped condensates \cite{Wolf2022,Jia2022}, fundamental tests \cite{Asenbaum2020} or studies on interspecies interactions \cite{DIncao2022,Chang2024,Dincao_2017}.

Consequently, it is critical for those experiments to realize an atom release where both species wouldn't acquire a differential velocity and remain co-located over the time of flight. 
To guarantee co-location during the free fall time, the transient differential accelerations $\Delta \boldsymbol{a}(t)$ must integrate to zero at the end of the release.
Assuming that the magnetic field remains constant over the size of the two overlapping BECs, $B(r_\mathrm{K},t)=B(r_{\mathrm{Rb}},t) \, \, \forall \, t$, the masses and accelerations of K and Rb are related through $a_\mathrm{Rb}(t)/a_\mathrm{K}(t)=m_\mathrm{K}/m_\mathrm{Rb}$. 
In the absence of oscillations of the two species at release, their corresponding co-location at release and after time-of-flight indicates a vanishing relative velocity. The latter has been confirmed by the simulation described in Sec.~\ref{sec:COMsimulation}. This model yields the evolution of trap minima and frequencies that is required for a comprehensive simulation of the respective density profiles after release.
\medskip
\begin{figure}
    \centering
    \includegraphics[width=0.95\textwidth]{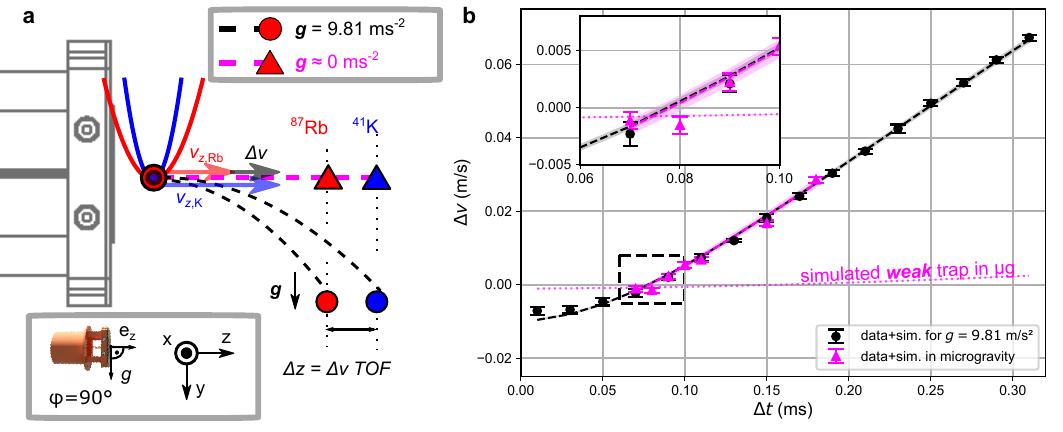}
    \caption{
    Measurement of the release kick and determination of an effectively force-free release of the atoms in the magnetic trap on ground for $\varphi=90^\circ$ and in microgravity.
    a) The atoms are released from the atom chip but acquire a differential initial velocity $\Delta v = v_{z, \mathrm{K}}-v_{z, \mathrm{Rb}}$ along $z$-direction due to the magnetic kick. Trapping potentials and atoms are shown in red ($^{87}$Rb) and blue ($^{41}$K), respectively.
    The red and blue arrows denote the release velocities of the two species due to the finite switch-off. The atomic trajectories are depicted by magenta (microgravity) and black (gravity, $\varphi=90^\circ$) dashed lines. To mitigate effects due to interspecies interactions, the experiments are conducted sequentially as single-species sequences.    
    b) By scanning the time delay $\Delta t$ between switching off the atom chip and the external coil, the magnitude of the release kick can be determined and minimized. The measurement values (black and magenta error bars) show the differential velocities $\Delta v$ between both ensembles. $\Delta v$ is extracted after a time-of-flight of 20.38$\,$ms (ground) and 40.34$\,$ms (microgravity), respectively. The dashed lines show the predictions based on a classical trajectory with time constants of $\tau_1=(31\pm3)\,$µs for the chip structures and $\tau_2=(925 \pm 46\,)$µs for the coils. The magenta dotted line is the expected switch-off kick in a weak magnetic trap used for atom interferometry in microgravity with eigenfrequencies $f_{\mathrm{Rb}}=\{11,34,31\}\,$Hz and $f_{\mathrm{K}}=\{16,50,45\}\,$Hz that is not accessible on ground. The shaded area shows the impact of a 5\% variation of the decay constants on the release velocities.
    The inset shows the behaviour of the measurement and simulations around the zero crossing. The remaining differential release velocity at $\Delta t=80\,$\textmu s is experimentally determined as $\Delta v = (-1.5\pm 0.7)\,$mm/s in a single point microgravity measurement. Note that the model suggests a smaller value of $(0.3\pm0.7)\,$mm/s. The reduced $\chi^2_r$ of the two-parameter fits amount to $\chi^2_r=1.15$ (ground model, black dashed line) and $\chi^2_r=3.60$ (microgravity model, magenta dashed line). The ground data represent the average of 10 consecutive measurements, while the microgravity data consist of single-shot measurements. Error bars are determined from the standard deviation of the atom cloud position on ground. Error bars for the µg case are extrapolated from the corresponding ground-based uncertainties and therefore do not include additional uncertainties introduced by the carrier platform.}
    \label{fig:ReleaseScheme}
\end{figure}

To minimize residual forces due to the release, we introduce a tunable delay $\Delta t$ between switching off the slow, high-inductance coils and the fast chip structures, while keeping the $B_x = -1.3\,$G offset field to preserve internal states (see methods section). By changing the value of $\Delta t$, it is possible to tune the magnetic forces sensed by the atoms during the switch-off transient.
In Fig.~\ref{fig:ReleaseScheme}, we show the extracted differential release velocities between the two clouds, determined by their relative positions in dependence of $\Delta t$ after a fixed time of flight. The release velocities obtained at a time of flight of $40.38\,$ms under microgravity conditions (magenta triangles) are consistent with the data recorded in the laboratory at a time of flight of $20.38\,$ms (black dots).
The zero crossing yields the ideal working point as it provides an effectively force-free release that allows to conduct unperturbed mixture studies. We point out that kicks along directions parallel to the atom chip are not observable. This is related to the condition that the position of the trap minimum depends solely on the currents through the chip wires $I_{SC}$ and $I_{BC}$, as well as the external bias field $B_y$, and is located along an axis perpendicular to both the chip surface and the bias field \cite{Folman2002}.
We model the release by solving the classical equations of motion of $^{41}$K and $^{87}$Rb in a time-dependent magnetic field (see methods Section~\ref{sec:COMsimulation}). The temporal evolution of this field is determined using a calibrated model of the atom chip and the external coils but does not account for the material surroundings. With a probe (Aim TTi Iprober 420) we observe that the coil current follows an approximately linear transient while the chip  structures reveal an exponential decay.
As eddy currents in the vacuum chamber and the atom chip mount delay in an unknown way the magnetic field change,  we introduce time constants $\tau_{1,2}$ for both field generating currents as free parameters without altering the general linear or exponential behavior.\\
The model successfully reproduces the observed data using an exponential decay with $\tau_1 = 31\,$\textmu s for the atom chip structures and a linear ramp of $\tau_2 = 925\,$\textmu s for the bias coils with a reduced $\chi^2=1.15$ (ground) and $\chi^2=3.60$ (microgravity). The higher $\chi^2$ in microgravity is likely due to an underestimation of the positional instability of the data points, which are single-shot measurements from consecutive experimental drops.
The final drift velocity $\Delta v = v_\mathrm{K} - v_{\mathrm{Rb}}$ between the two species at the optimal point depends on the switch-off time with a linear dependency of $\delta \Delta v / \delta \Delta t = (272\pm 16)\,$\textmu m/ms$^2$.\\
To estimate the differential kick under microgravity conditions, we simulate the release from a weak trap, ideally suited for dual-species equivalence principle tests in microgravity (dotted magenta curve in Fig.~\ref{fig:ReleaseScheme}b). Due to the lower magnetic forces of weak traps, the effect is naturally suppressed by an order of magnitude and amounts to $\delta \Delta v / \delta \Delta t = (11.8\pm 0.7)\,$\textmu m/ms$^2$. With an assumed time-resolution of $100\,$ns of a modern control system this would result in $\Delta v = 1.2\,$\textmu m/s which is far below previously reported values \cite{Modugno2000, Gaaloul2022} and meets the requirements for tests of the equivalence principle with Eötvös parameters in the range of $10^{-15}$ \cite{PhysRevD.102.124043}.

\subsection{Interacting quantum mixtures}
\label{sec:interaction}

\begin{figure}
    \centering
    \includegraphics[width=1.0\textwidth]{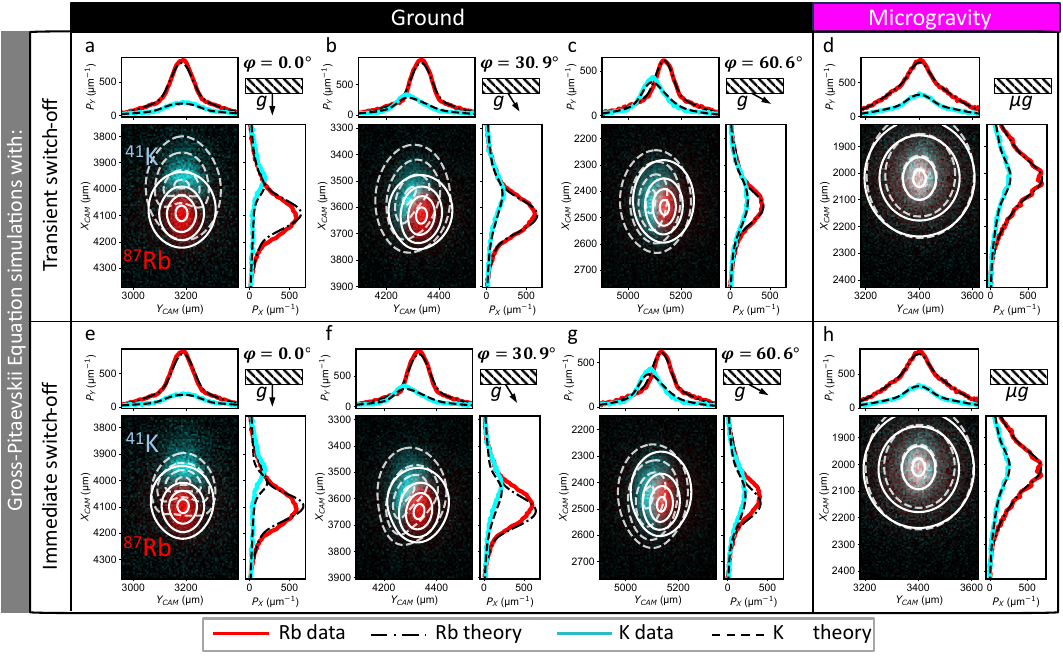}
    \caption{Absorption images of mixed BECs of $^{41}$K and $^{87}$Rb after free expansion, measured at different apparatus orientations ($\varphi=0^\circ$, $30.9^\circ$, and $60.6^\circ$) and in microgravity. The angle of the gravity vector with respect to the atom chip is depicted in the upper right corner. The horizontal and vertical plots show the projected 1D line densities, with solid lines representing experimental data (cyan for $^{41}$K, red for $^{87}$Rb) and black lines for simulation results (dashed for $^{41}$K, dash-dotted for $^{87}$Rb). Panels a-d show simulations with the full model including a transient magnetic trap switch-off, and e-h show simulations with an instantaneous switch-off. Atom numbers for the BECs and thermal fractions were determined by fitting the line densities with Gaussian and Thomas-Fermi functions. Atom numbers in BEC fraction: $N_\mathrm{K} = 1.44\cdot10^4$, $N_{\mathrm{Rb}} = 4.39\cdot10^4$ at $\varphi = 0^\circ$; $N_\mathrm{K} = 2.02\cdot10^4$, $N_{\mathrm{Rb}} = 4.69\cdot10^4$ at $\varphi = 30.9^\circ$;  $N_\mathrm{K} = 2.32\cdot10^4$, $N_{\mathrm{Rb}} = 2.77\cdot10^4$ at $\varphi = 60.6^\circ$; and $N_\mathrm{K} = 1.78\cdot10^4$, $N_{\mathrm{Rb}} = 1.31\cdot10^4$ in microgravity.} 
    \label{fig:RotatedSetup}
\end{figure}

Using the above described methods to find a suitable switch-off protocol, we study interacting quantum mixtures in presence of the Earth's gravitational field and compare the results with those obtained under microgravity. While being partially suppressed, the system remains sensitive to the shutdown of the magnetic trapping potentials, as illustrated in this section. This gives rise to a complex interplay between gravity, interatomic repulsion and magnetic release forces. Having a solid understanding of this interplay is an important step towards future space-borne experiments with high demands on absolute or differential release velocities \cite{Ahlers2022, leveque2022carioqa, Lundblad2023}.
In addition to microgravity experiments, the inherent portability of our experiment offers the unique possibility to adjust the orientation of the experimental chamber (and so the atom chip) with respect to gravity. For that purpose, we mount the physics package into a rotatable frame which defines the rotation axis as shown in Fig.~\ref{fig:setup} b,c). 
In this way we disentangle the forces due to
the decaying magnetic field, which mainly acts perpendicular to the atom chip surface, from the forces due to gravity and due to the mutual repulsion of the gases. The latter is aligned with gravity as the two gases being located one on top of the other. In microgravity, the absence of the differential gravitational sag modifies the ground state. As a result, the mixture does not phase separate along the direction of gravity but still remains compatible with both symmetric and antisymmetric (side-by-side) configurations \cite{Wolf2022, Elliott2023,Pichery2023}.\\
In Fig.~\ref{fig:RotatedSetup}a-c, we show absorption images of $^{41}$K (cyan) and $^{87}$Rb (red) BEC mixtures released from a magnetic trap for three different angles to gravity. In all cases, the $^{41}$K BECs are well separated from the $^{87}$Rb BECs and aligned to the gravity vector. In our release trap, the gravitational sag-induced displacement of the trap minima is only on the order of 1$\,$µm. Thus, the center-of-mass separation of both species is primarily due to the repulsive interaction, as both condensates align in the trap along gravity, thereby increasing the shift of the clouds in this direction. Fig.~\ref{fig:RotatedSetup}d shows the microgravity outcome, with the two atomic clouds overlapping after the time-of-flight, strikingly distinct from the behavior observed in the presence of gravity.\\
We perform a comprehensive simulation of the ground state and evolution of the BEC mixture taking into account the previously described dynamics of the magnetic trap release. The simulation is based on a numerical toolkit solving the 3D time-dependent Gross-Pitaevskii equations \cite{Pichery2023, Seckmeyer2025} and has been extended for time-varying potentials to capture the magnetic field transients.\\
The calculated density profiles are shown in Fig.~\ref{fig:RotatedSetup}a-d on top of the experimental data. Although the mechanical mount allows to tune the angle in steps of $\Delta \varphi=2.5^\circ$, we find the exact angle by evaluating the absolute $x$ and $y$ position of the $^{87}$Rb cloud after 25$\,$ms time of flight to account for possible deviations. 
Apart from the refinement of the angle, the simulation does not contain any free parameters and shows the expected absolute positions of the atomic densities projected to the detection system axis (see sec.~\ref{sec:MixModel} for more details).\\
The simulated density profiles in microgravity and for rotation angles of $30.9^\circ$ and $60.6^\circ$ show excellent agreement with the experimental data in positions and shapes. For the $0.0^\circ$ configuration (Fig.~\ref{fig:RotatedSetup}a), a deviation of 57$\,\textrm{\textmu m}\pm10\,\textrm{\textmu m}$(exp) between the relative distances of the two density maxima remains between simulation and experiment. To quantify the impact of the decaying magnetic fields on the atoms, we compare our simulated density distributions with a second simulation assuming an immediate release without decaying magnetic fields, shown in Fig.~\ref{fig:RotatedSetup} (e-h). By comparing the calculated positions of the atoms in (a-d) with the respective positions in (e-h) we can directly quantify the impact of the transient magnetic fields. For all rotation angles, the decaying magnetic fields are acting in the direction normal to the atom chip surface while the interaction driven acceleration is directed along the separation due to the differential gravitational sag. The positions horizontally to the atom chip are not affected by the release. For $\varphi=0.0^{\circ}$, the transient magnetic fields account for $18\,\textrm{\textmu m}$ of the observed separation of the two species.\\ 
We also consider the impact of background magnetic field gradients along the direction of free fall and systematic errors in the estimation of absolute atom numbers which would impact the simulated repulsion. From time of flight measurements of differential mixture positions we obtain an upper limit for a residual magnetic field gradient of $B_z'<0.025\,$G/cm which leads to an additional relative separation of $<5.6\,\textrm{\textmu m}$. By increasing the simulated Rb (K) atom number by $10\%$, the differential position changes by $5.8\,\textrm{\textmu m}$ ($-2.3\,\textrm{\textmu m}$).
Given the magnitude of the different systematic effects, we cannot attribute the observed deviations in the 0.0$^\circ$ case to additional background fields or atom number uncertainties. Instead, they can be explained by the trajectories of the two BECs. During release, their separation temporarily reduces along $z$ due to transient magnetic gradients. Only in the 0.0$^\circ$ case, the associated atomic motion co-aligns with gravity and their mutual repulsion. This increases the interatomic repulsion compared to non-zero rotation angles. Therefore, we attribute the deviation in the 0.0$^\circ$ case to an oversimplified release model that does not capture the full details of the atomic motion during the transient.
We emphasize that the observed inconsistency in absolute position is below the size of the expanded BECs and only present in the $0.0^\circ$ configuration for Rb.
\clearpage

\section{Discussion}
We have developed a fully integrated sounding rocket payload for a two-species quantum gas experiment which is able to produce BEC mixtures of $^{87}$Rb and $^{41}$K with up to $2.5\cdot 10^5$ ($^{87}$Rb) and $5\cdot 10^4$ ($^{41}$K) atoms in a single species optimized sequence. By tuning the duration of the $^{87}$Rb-MOT we are able to tune the ratio of the BEC mixtures, allowing to work with $2\cdot10^4$ atoms in each condensate with a short preparation time of 2.3$\,$s. This represents, to the best of our knowledge, the highest flux of BEC mixture generation.\\
To further increase the atom numbers in the context of future projects, it will be necessary to increase the number of initially trapped $^{41}$K atoms in the magnetic trap but also to further limit losses due to inelastic collisions between the species during evaporation. Improvements in laser cooling of $^{41}$K, for example by D1-cooling techniques, could significantly increase the number of magnetically trapped atoms \cite{Chen2016}. Inelastic collisions can be further reduced by increasing the power of the depumper microwave. Techniques such as optical shielding have also been suggested as an effective way to reduce inelastic collisions in ultracold atomic mixtures\cite{Xie2022}.

We have shown the impact of realistic magnetic field decays of the external coils and chip structures and observed a differential release velocity between both species. By introducing a time delay between switching off the magnetic coil and the atom chip we were able to reduce the differential release velocity. The technique is compatible with requirements on the differential velocity of future high-precision measurements such as the tests of the Einstein equivalence principle. \\
We note that the spatial evolution of expanding BEC mixtures in different interaction regimes has been studied in a number of experiments \cite{Wang2016,Wacker2015,Papp2008,Lee2018}. In these realizations, atoms were trapped in optical dipole traps and are naturally not affected by magnetic forces during release. Although recent progress has demonstrated the feasibility of dipole traps in microgravity \cite{Vogt2020PRA, haase2025robustcompactsinglelenscrossedbeam}, atom chips are widely used in compact experiments and represent the tool of choice in environments with high demands on robustness and low power consumption.\\
We could demonstrate the robustness of the system and the sequences by rotating the payload with respect to gravity between 0 and 90$^\circ$ and generated BEC mixtures in each configuration, between 0 and 75$^\circ$ without changing sequence parameters. The rotation allowed us to study the different forces acting on the atoms, in particular gravitational forces, interspecies repulsion and magnetic forces generated by eddy currents during release.\\ By applying our switch-off protocol in microgravity with this payload, the dynamics of the atoms are only governed by their intra- and interspecies interactions. In absence of the gravitational sag, the mixture ground state can show various geometric configurations, such as symmetric splitting of the $^{41}$K-atoms in cigar-shaped traps \cite{Pichery2023} or closed shells of $^{41}$K forming bubble traps~\cite{Wolf2022}. The interaction driven expansion in microgravity depends on the details of the underlying ground state geometry and will be studied in future campaigns in the Einstein-Elevator. Our findings offer therefore a detailed understanding of the release processes and minimization of additional transient forces that are necessary for future quantum mixture experiments relevant for many-body physics as well as for quantum sensing.\\ 
In conclusion, we are able to generate BEC mixtures of $^{87}$Rb and $^{41}$K with a high flux in different orientations to gravity as well as in microgravity. By introducing a suitable time delay between switching off the atom chip and the coils, we can tune the differential release velocity to zero, only limited by the time resolution of the hardware. The effects of gravity, interactions and transient magnetic fields are well described by a set of coupled Gross-Pitaevskii equations. Our results lay the foundation for further studies with BEC mixtures in microgravity, enabling a wide range of proposed experiments ranging from tests of the equivalence principle \cite{Ahlers2022}, the generation of quantum droplets \cite{Petrov2015,Derrico2019} to applications in molecular physics \cite{Dincao_2017, DIncao2022}.

\section{Methods}
\subsection{Apparatus design and sequence details}
%
All experiments reported here were performed on ground with the sounding rocket payload MAIUS-B, designed for the generation of $^{41}$K and $^{87}$Rb BECs in space. The payload was launched within the sounding rocket flight campaign MAIUS-2 on 2nd of December 2023. Details on the overall payload design can be found in \cite{Elsen2023}.

\begin{figure}
    \centering
    \includegraphics[width=1\textwidth]{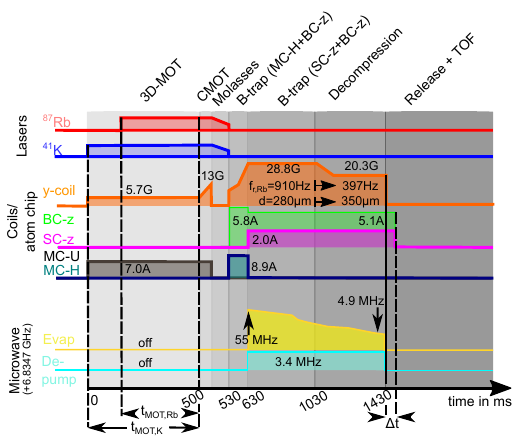}
    \caption{Experimental sequence for BEC mixture generation and release studies (BC: Base-Chip, SC: Science-Chip, MC: Mesoscopic chip). The number ratio is tuned by changing the duration $t_{\textrm{MOT,Rb}}$. The release is analyzed and optimized by scanning $\Delta t$. $f_{r,Rb}$ is the radial frequency of the $^{87}$Rb trap, $d$ its distance to the atom chip surface. The time axis is not to scale. Further details of the experimental sequence can be found in \cite{Piest2021}.}
    \label{fig:sequence}
\end{figure}

The BEC generation is performed in a two chamber design, consisting of a source and a science chamber. The source chamber generates a cold, dual-species atomic beam by overlapping a $^{41}$K and a $^{87}$Rb 2D-MOT. The pusher laser beam guides the cold atoms through a differential pumping stage towards the science chamber, where they are trapped by a three-dimensional mirror MOT, provided by four laser beams, an atom chip and three pairs of Helmholtz coils along all spatial directions. In a subsequent step, the atom chip Z-structures form the Ioffe-Pritchard trap together with a bias field generated by the y-coils for magnetic confinement of $^{41}$K and a $^{87}$Rb (see Fig.~\ref{fig:sequence}). At this stage, evaporative cooling of $^{87}$Rb is performed with two microwave frequencies that are coupled via U-shaped antennas on the base chip layer. The atom clouds are imaged in consecutive experimental runs using absorption imaging \cite{Reinaudi2007} with two lenses in $F-2F-F$ configuration ($F=50\,$mm) and circularly polarized light. After a variable time of flight, the atoms are imaged with a 20$\,$µs imaging pulse resonant to the $\ket{S_{1/2}, F=2} \rightarrow \ket{P_{3/2}, F=3}$ transition of $^{41}$K or $^{87}$Rb, respectively. 

Before integration of the full payload, the cold atom apparatus was mounted in a rotatable aluminium frame to characterize the influence of different gravity directions. This allowed a rotation around the weak (longitudinal) trapping axis of the cylindrical trap in steps of 2.5°, resulting in relative change of the direction of gravity with respect to the magnetic trap. This way, gravity never points along the weak axis which makes it possible to trap the atoms for all angles.

\subsection{Atom losses and collisions}
\label{sec:losses}
\begin{figure}[tb]
    \centering
    \includegraphics[width=1.0\textwidth]{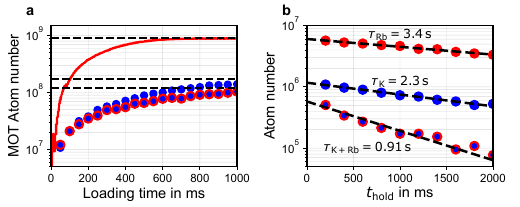}
    \caption{a) MOT loading of $^{87}$Rb (red), $^{41}$K (blue), and $^{41}$K with simultaneously loaded $^{87}$Rb atoms (red+blue). The saturation values are indicated with black-dashed lines ($N_{\textrm{Rb}}=9.0\cdot10^8, N_{\textrm{K}}=1.7\cdot10^8, N_{\textrm{K+Rb}}=1.2\cdot10^8$). The number of $^{41}$K atoms in the combined MOT decreases by a factor of 0.29. b) Lifetime measurement of  magnetically trapped $^{87}$Rb (red), $^{41}$K (blue), and $^{41}$K with simultaneously trapped $^{87}$Rb atoms (red+blue). The measurements were performed at a temperature of 100$\,$\textmu K and in a trap with frequencies $f_{x, \mathrm{Rb}}=23\,$Hz and $f_{r, \mathrm{Rb}}=910\,$Hz for $^{87}$Rb, and $f_{x, \mathrm{K}}=33\,$Hz and $f_{r, \mathrm{K}}=1326\,$Hz for $^{41}$K.}
    \label{fig:KRbInfluence}
\end{figure}

There are different mechanisms that lead to a decrease of $^{41}$K atoms in the BEC with increasing number of $^{87}$Rb atoms trapped initially in the MOT. If loaded simultaneously in the MOT, collisions between $^{41}$K and $^{87}$Rb lead to losses within the $^{41}$K-MOT. As shown in Fig.~\ref{fig:KRbInfluence}a we see an impeding effect of the $^{87}$Rb-MOT to the $^{41}$K-MOT.
In the magnetic trap, the most important loss process is inelastic collisions between $^{41}$K atoms in the state $\ket{F=2, m_F=2}$ and $^{87}$Rb atoms in the impurity state $\ket{F=2, m_F=1}$ \cite{Fer02}. Detrimental effects of remaining $^{87}$Rb atoms in $\ket{F=2, m_F=1}$ were observed previously in other setups \cite{Burchianti2018, Wack18}.
Fig.~\ref{fig:KRbInfluence}b shows the comparison of the different lifetimes of magnetically trapped $^{87}$Rb atoms (3.5$\,$s), $^{41}$K atoms (2.3$\,$s) and $^{41}$K in presence of $4.7\cdot10^7$ $^{87}$Rb atoms (0.91$\,$s). By switching on the depumper microwave which is used to depump co-trapped $^{87}$Rb atoms in the impurity state $\ket{F=2,m_f=1}$, the lifetime of $^{41}$K increases from 0.91$\,$s to 1.4$\,$s. We attribute the remaining difference in lifetime to $^{87}$Rb atoms in the state $\ket{F=2, m_F=1}$ getting continuously repumped from the anti-trapped state $\ket{F=1, m_F=1}$ by the evaporation microwave after they moved by some distance. 

\subsection{Simulation of center-of-mass dynamics and trap configurations}
\label{sec:COMsimulation}
The center-of-mass motion of the released atoms, which models the trap release discussed in Section~\ref{sec:releasedynamics}, is determined by solving the classical equations of motion. These include the time-varying potential generated by the magnetic field $\textbf{B}(\textbf{r},t)$. For a given time $t$, the magnetic field configuration is calculated by adding up the magnetic fields of the chip structures and coils using the Biot-Savart law
\begin{equation}
    \textbf{B}(\textbf{r},t) = \sum_j \textbf{B}_j(\textbf{r},t) = \frac{\mu_0}{4\pi} \sum_j I_j(t) \oint_{C_j} \frac{d\textbf{l}_j \times \textbf{r'}_j}{r'^3_j},
\end{equation}
with the time-dependent currents $I_j(t)$ of the atom chip structures and external coils, the magnetic constant $\mu_0$, the line element $d\textbf{l}_j$ along the structure $C_j$ and the distance $\textbf{r'}_j$ between the line element $d\textbf{l}_j$ and $\textbf{r}$. We calibrated our magnetic field model by evaluating positions of trap minima and their respective magnetic field magnitudes and trap frequencies over a wide range of relevant traps. As gauging parameters, we used scaling factors for the X- and Y-coil bias fields, constant magnetic field offsets $B_x, B_y$ and z-direction shifts of the science and base chip layers.
The residual deviation between model predictions and experimental measurements in the trap frequencies and positions is below $2\%$ across the range of traps studied in the Results section. \\
The potential is given by the Zeeman energy shift
\begin{equation}
    U(\textbf{r},t)=\mu_B \, g_F \, m_F \, B(\textbf{r},t)
\end{equation}
with $B(\textbf{r},t)$ being the modulus of the magnetic field $\textbf{B}(\textbf{r},t)$, $\mu_B$ the Bohr magneton, $g_F=1/2$ the Landé g-factor and $m_F=2$ the magnetic quantum number.
The equations of motion for the $^{87}$Rb and $^{41}$K atoms are given by:
\begin{equation}
    \ddot{q_j}=-\frac{1}{m}\frac{\partial U(\textbf{r},t)}{\partial q_j},
\end{equation}
where $q_j \in \left\{x,y,z\right\}$ and $m$ the respective atomic mass.
These equations are respecting the initial conditions $q_j(0)= q_{j,0}$ and $\dot q_j(0) = \ddot q_j(0)=0$ for both species where $q_{j,0}$ is the trap minimum at $t = 0$.
To solve the equations of motion, we use the leap-frog integration method \cite{Tuckerman2010} with time steps of $\delta t=10\,$\textmu s.\\
{The chip wires exponentially decay from an initial current $I_0$ to a final current $I_1$ with a time constant of $\tau_1 = 31\,$\textmu s:
\begin{equation}
    I_\mathrm{w}(t) = (I_0 - I_1) \, e^{-t/\tau_1} + I_1.
\end{equation}

The current in the y-coils is modeled by a linear ramp moving from $I_0$ to $I_1$ in $\tau_2 = 925\,$\textmu s.
Therefore, they are described by:
\begin{equation}
    I_\mathrm{c}(t) = \left( \frac{I_0 - I_1}{\tau_2} t + I_0 \right) \, \Theta(\tau_2 - t) + I_1 \,  \Theta(t - \tau_2),
\end{equation}
where $\Theta(t)$ is the heaviside function defined as:
\begin{equation}
    \Theta(t) = 
    \left\{
        \begin{aligned}
            1 & \text{ if $t \geq 0$} \\
            0 & \text{ if $t < 0$}
        \end{aligned}
    \right. \, .
\end{equation}
}

\subsection{3D-modelling of BEC mixture}
\label{sec:MixModel}

The simulation of the absorption images is based on a three-step process: calculation of the BEC ground state, propagation during free-fall and finally, projection onto the imaging plane. The ground state is determined using the imaginary time propagation method to solve the coupled Gross-Pitaevskii equation (GPE) describing an interacting BEC mixture \cite{Bao2004,Lehtovaara2007}. The s-wave scattering lengths used to calculate the interspecies and intraspecies interactions are $a_{\mathrm{Rb}}=98.96 \, a_0$ \cite{Marte2002}, $a_\mathrm{K}= 60.54 \, a_0$ \cite{Falke2008} and $a_{\mathrm{KRb}}=165.3 \, a_0$ \cite{Ferlaino2006} in units of the Bohr radii $a_0$. The trapping potential is approximated harmonically around the potential minimum using the magnetic field from the chip model. This model provides the characteristics of the trapping potential such as the minimum position, frequencies and orientation of the eigenvectors. In general, minimum positions and eigenvectors are different for both species $^{41}$K and $^{87}$Rb due to the gravitational sag. The gravity vector is set according to the tunable orientation of the apparatus. The atom numbers of the simulated ensembles are determined experimentally using absorption imaging of the two clouds \cite{Reinaudi2007}.\\
The ground state is then used as the initial state to simulate the dynamics of the mixture during its free expansion by solving the coupled GPE with the scaled grid method presented in \cite{Pichery2023, Seckmeyer2025}. This enables an efficient calculation of the dynamics as the simulation volume increases by $10^4$ during the 25.38 ms expansion. The CPU time required for the free expansion dynamics is roughly 50 minutes of CPU time, and 4 minutes of real time on a cluster using 16 cores in parallel, for a spatial grid of $(N_x=64,N_y=64,N_z=128)$ points.\\
The evolution of the BEC mixture is described in the coordinates of the atom chip (see Fig.~\ref{fig:ReleaseScheme}). During the switch-off ramp, the trap characteristics (frequencies and position) are extracted from the chip model by evaluating the Hessian matrix of the minimum position. However, within this sequence, this estimate shows divergences over very short time intervals (less than 10 µs), due to a brief and transient dominance of potential anharmonicities. Given their extremely short duration, these quasi-instantaneous irregularities are smoothed out, which enables agreement between simulation and experimental results.
The free expansion of the atoms is considered as a free fall in the calculations, the final position of the atoms takes into account the distance travelled by the atoms during the time of flight. The chip edge is measured on the camera and calibrates the origin of the coordinate frame.
To visualize the atoms in the camera frame, we integrate the 3D density along the optical axis of the detection system $\textbf{n}=(\cos(\theta),\sin(\theta),0)$ with $\theta = 46.1^\circ$.
To take into account the finite resolution of the detection system, the acquired density images are convoluted with a Gaussian distribution of width $\sigma = 15$ \textmu m.  

For better comparability, the thermal fractions of both ensembles are extracted from the Gaussian part of a bimodal fit from the absorption images and added to the simulated data, using the same method as in \cite{Pichery2023,Elliott2023}.

\section{Data availability}
The data generated in this study, in particular to generate Fig. 2, 4, 5 and 7 are available on the  figshare repository under \url{https://doi.org/10.6084/m9.figshare.31370050}.

\section{Code availability}
Two-species codes are based on FFT propagation examples published in reference \cite{Seckmeyer2025} optimized by the methods published in \cite{Pichery2023}. Simulations presented in Fig. 4b are based on a non-disclosed, experiment-specific, gauged atom-chip model.

\bibliography{References}
\section{Acknowledgements}
The QUANTUS IV - MAIUS project is a collaboration of Zentrum für angewandte Raumfahrttechnologie und Mikrogravitation Bremen, Leibniz Universität Hannover, Humboldt-Universität zu Berlin, Johannes Gutenberg-Universität Mainz and Ferdinand-Braun-Institut, Leibniz-Institut für Höchstfrequenztechnik. We acknowledge support from Deutsches Zentrum für Luft- und Raumfahrt - Raumfahrtbetrieb, Oberpfaffenhofen, Deutsches Zentrum für Luft- und Raumfahrt - Simulations- und Softwaretechnik, Braunschweig. P.G., D.L. and J.B. would like to thank Sebastian Lazar and Moritz Möbius for their support and operation of the Einstein-Elevator.

\section{Funding}
This work is supported by the German Space Agency DLR with funds provided by the Federal Ministry for economic affairs and climate action (BMWK) under grant number DLR 50WP 1431-1435. T.E. and G.M. acknowledge support from DLR grants 50WM2245-A (CAL-II), 50WM2545A (CAL-III). We acknowledge support by the Deutsche Forschungsgemeinschaft (DFG, German Research Foundation) under Germany’s Excellence Strategy - EXC-2123 QuantumFrontiers - 390837967. 

\section{Author contributions}
B.P, J.B., P.G. and D.L. conducted the experiments under supervision of E.M.R. P.G., D.L. and J.B. conducted the microgravity campaigns in Einstein-Elevator under supervision of C.L. and E.M.R. Data analysis and interpretation were performed by B.P., J.B., T.E., A.Pi., P.G., N.G. and E.M.R. Numerical simulations were provided by T.E., A.Pi. and G.M. under supervision of N.G. and E.C.
The development and commissioning of the apparatus was led by J.G. with contributions from B.P, J.B., P.A., W.B., S.B., K.D., M.E., O.H., C.K., A.Pa., C.R., A.We. and T.W.\\
B.P, J.B., T.E. and A.Pi. wrote the manuscript with feedback from all authors. E.C., N.G., J.G., C.L, A.Pe. K.S., A.Wi, P.W. and E.M.R. are principal investigators of the contributing groups. J.G. is the principal investigator of the project. Correspondence to E. M. Rasel (rasel@iqo.uni-hannover.de) and B. Piest (baptist.piest@obspm.fr)

\section{Competing interests}
The authors declare no competing interests.

\section{Figure Legends}
\subsection{Figure 1}
Overview of the experimental setup. a) Overall payload. The physics package is highlighted with a white dashed box. OBC: on-board computer. b) 3D drawing of the MAIUS-B physics package. When fixed in a suitable mounting frame, the module is rotatable around the indicated rotation axis, here shown for $\varphi=0^\circ$. c) Orientation of the atom chip inside the vacuum chamber for three different angles $\varphi$ between $z$-axis and gravity direction $g$. Fig. 1a) is adapted from \cite{Elsen2023}, licensed under CC BY 4.0. Modifications include revised annotations and rotation of image.
\subsection{Figure 2}
Generation of BEC mixtures with different ratios in the BEC fraction of the two species on ground. Corresponding absorption images for four different configurations are shown exemplarily on the right. The images are normalized for each species separately for better visibility. The inset shows typical loading curves of the magnetic traps. In this example, we load the K-MOT for 500$\,$ms and the Rb-MOT for 300$\,$ms. The possible magnetic trap loading curves for longer times is highlighted in grey.
\subsection{Figure 3}
Comparison of atom numbers and preparation times of BECs of $^{87}$Rb and $^{41}$K in various experiments. The red (blue) dots represent the atom number in the BEC fractions of $^{87}$Rb ($^{41}$K). The dotted lines indicate the constant atomic flux (numbers on the left side). The required capabilities of the future space-borne experiments BECCAL and STE-QUEST are also plotted. Note that the data points show the atom numbers for the single-species-optimized operation except for STE-QUEST, where they are targeted in mixture operation. Sources: \cite{Burchianti2018,Elliott2018,Thalhammer2008,Modugno2001,Elliott2023,Frye2021,Ahlers2022}.
\subsection{Figure 4}
Measurement of the release kick and determination of an effectively force-free release of the atoms in the magnetic trap on ground for $\varphi=90^\circ$ and in microgravity.
    a) The atoms are released from the atom chip but acquire a differential initial velocity $\Delta v = v_{z, \mathrm{K}}-v_{z, \mathrm{Rb}}$ along $z$-direction due to the magnetic kick. Trapping potentials and atoms are shown in red ($^{87}$Rb) and blue ($^{41}$K), respectively.
    The red and blue arrows denote the release velocities of the two species due to the finite switch-off. The atomic trajectories are depicted by magenta (microgravity) and black (gravity, $\varphi=90^\circ$) dashed lines. To mitigate effects due to interspecies interactions, the experiments are conducted sequentially as single-species sequences.    
    b) By scanning the time delay $\Delta t$ between switching off the atom chip and the external coil, the magnitude of the release kick can be determined and minimized. The measurement values (black and magenta error bars) show the differential velocities $\Delta v$ between both ensembles. $\Delta v$ is extracted after a time-of-flight of 20.38$\,$ms (ground) and 40.34$\,$ms (microgravity), respectively. The dashed lines show the predictions based on a classical trajectory with time constants of $\tau_1=(31\pm3)\,$µs for the chip structures and $\tau_2=(925 \pm 46\,)$µs for the coils. The magenta dotted line is the expected switch-off kick in a weak magnetic trap used for atom interferometry in microgravity with eigenfrequencies $f_{\mathrm{Rb}}=\{11,34,31\}\,$Hz and $f_{\mathrm{K}}=\{16,50,45\}\,$Hz that is not accessible on ground. The shaded area shows the impact of a 5\% variation of the decay constants on the release velocities.
    The inset shows the behaviour of the measurement and simulations around the zero crossing. The remaining differential release velocity at $\Delta t=80\,$\textmu s is experimentally determined as $\Delta v = (-1.5\pm 0.7)\,$mm/s in a single point microgravity measurement. Note that the model suggests a smaller value of $(0.3\pm0.7)\,$mm/s. The reduced $\chi^2_r$ of the two-parameter fits amount to $\chi^2_r=1.15$ (ground model, black dashed line) and $\chi^2_r=3.60$ (microgravity model, magenta dashed line). The ground data represent the average of 10 consecutive measurements, while the microgravity data consist of single-shot measurements. Error bars are determined from the standard deviation of the atom cloud position on ground. Error bars for the µg case are extrapolated from the corresponding ground-based uncertainties and therefore do not include additional uncertainties introduced by the carrier platform.
\subsection{Figure 5}
Absorption images of mixed BECs of $^{41}$K and $^{87}$Rb after free expansion, measured at different apparatus orientations ($\varphi=0^\circ$, $30.9^\circ$, and $60.6^\circ$) and in microgravity. The angle of the gravity vector with respect to the atom chip is depicted in the upper right corner. The horizontal and vertical plots show the projected 1D line densities, with solid lines representing experimental data (cyan for $^{41}$K, red for $^{87}$Rb) and black lines for simulation results (dashed for $^{41}$K, dash-dotted for $^{87}$Rb). Panels a-d show simulations with the full model including a transient magnetic trap switch-off, and e-h show simulations with an instantaneous switch-off. Atom numbers for the BECs and thermal fractions were determined by fitting the line densities with Gaussian and Thomas-Fermi functions. Atom numbers in BEC fraction: $N_\mathrm{K} = 1.44\cdot10^4$, $N_{\mathrm{Rb}} = 4.39\cdot10^4$ at $\varphi = 0^\circ$; $N_\mathrm{K} = 2.02\cdot10^4$, $N_{\mathrm{Rb}} = 4.69\cdot10^4$ at $\varphi = 30.9^\circ$;  $N_\mathrm{K} = 2.32\cdot10^4$, $N_{\mathrm{Rb}} = 2.77\cdot10^4$ at $\varphi = 60.6^\circ$; and $N_\mathrm{K} = 1.78\cdot10^4$, $N_{\mathrm{Rb}} = 1.31\cdot10^4$ in microgravity.
\subsection{Figure 6}
Experimental sequence for BEC mixture generation and release studies (BC: Base-Chip, SC: Science-Chip, MC: Mesoscopic chip). The number ratio is tuned by changing the duration $t_{\textrm{MOT,Rb}}$. The release is analyzed and optimized by scanning $\Delta t$. $f_{r,Rb}$ is the radial frequency of the $^{87}$Rb trap, $d$ its distance to the atom chip surface. The time axis is not to scale. Further details of the experimental sequence can be found in \cite{Piest2021}.
\subsection{Figure 7}
a) MOT loading of $^{87}$Rb (red), $^{41}$K (blue), and $^{41}$K with simultaneously loaded $^{87}$Rb atoms (red+blue). The saturation values are indicated with black-dashed lines ($N_{\textrm{Rb}}=9.0\cdot10^8, N_{\textrm{K}}=1.7\cdot10^8, N_{\textrm{K+Rb}}=1.2\cdot10^8$). The number of $^{41}$K atoms in the combined MOT decreases by a factor of 0.29. b) Lifetime measurement of  magnetically trapped $^{87}$Rb (red), $^{41}$K (blue), and $^{41}$K with simultaneously trapped $^{87}$Rb atoms (red+blue). The measurements were performed at a temperature of 100$\,$\textmu K and in a trap with frequencies $f_{x, \mathrm{Rb}}=23\,$Hz and $f_{r, \mathrm{Rb}}=910\,$Hz for $^{87}$Rb, and $f_{x, \mathrm{K}}=33\,$Hz and $f_{r, \mathrm{K}}=1326\,$Hz for $^{41}$K.
\end{document}